\documentclass[useAMS,usenatbib]{mn2e}

\usepackage{graphicx} 
\usepackage{rotating}
\usepackage{upgreek}

\title[Witnessing the Emergence of a Carbon Star]{Witnessing the Emergence of a Carbon Star}
\author[L. Guzman-Ramirez, et al.]{L. Guzman-Ramirez$^{1}$\thanks{Email:guzmanl@eso.org}, E. Lagadec$^2$, R. Wesson$^1$, A.~A. Zijlstra$^3$, A. M\"uller$^1$,\newauthor D. Jones$^{4,5}$, H.~M.~J. Boffin$^1$, G.~C. Sloan$^6$, M.~P. Redman$^7$,  A. Smette$^1$, \newauthor A.~I. Karakas$^{8,9}$ \& Lars-$\mathring{\rm A}$ke Nyman$^{1,10}$\\
$^{1}$European Southern Observatory, Alonso de C\'ordova 3107, Vitacura, Santiago, Chile \\
$^{2}$Laboratoire Lagrange, Universit\'e de Nice Sophia-Antipolis, Observatoire de la C$\hat{\rm o}$te d'Azur, CNRS, 06304 Nice, France \\
$^{3}$Jodrell Bank Centre for Astrophysics, School of Physics and Astronomy, University of Manchester, Manchester, M13 9PL, UK\\
$^{4}$Instituto de Astrof\'isica de Canarias, 38200 La Laguna, Tenerife, Spain \\
$^{5}$Departamento de Astrof\'isica, Universidad de La Laguna, Tenerife, Spain\\
$^{6}$Astronomy Department, Cornell University, Ithaca, NY 14853-6801, USA \\
$^{7}$Centre for Astronomy, School of Physics, National University of Ireland Galway, Galway, Ireland \\
$^{8}$Research School of Astronomy \& Astrophysics, Mt Stromlo Observatory, Australian National University, Canberra, Australia \\
$^{9}$Kavli IPMU (WPI), The University of Tokyo, Tokyo, Japan\\
$^{10}$Joint ALMA Observatory, Alonso de C\'ordova 3107, Vitacura, Santiago, Chile}
\begin{document}

\date{Accepted, 9 April 2015. Received, 8 April 2015; in original form, 16 January 2015}

\pagerange{\pageref{firstpage}--\pageref{lastpage}} \pubyear{2015}

\maketitle

\label{firstpage}

\begin{abstract}
During the late stages of their evolution, Sun-like stars bring the products of
nuclear burning to the surface. Most of the carbon in the Universe is believed to originate from stars with masses up to a few solar masses. Although there is a
chemical dichotomy between oxygen-rich and carbon-rich evolved stars, the
dredge-up itself has never been directly observed.  In the last three decades,
however, a few stars have been shown to display both carbon- and oxygen-rich
material in their circumstellar envelopes. Two models have been proposed to
explain this dual chemistry: one postulates that a recent dredge-up of carbon
produced by nucleosynthesis inside the star during the Asymptotic Giant Branch
changed the surface chemistry of the star. The other model postulates that oxygen-rich material exists in stable keplerian rotation around the central
star. The two models make contradictory, testable, predictions on the location
of the oxygen-rich material, either located further from the star than the
carbon-rich gas, or very close to the star in a stable disk. Using the Faint
Object InfraRed CAmera (FORCAST) instrument on board the Stratospheric
Observatory for Infrared Astronomy (SOFIA) Telescope, we obtained images of
the carbon-rich planetary nebula (PN) BD\,+30$^{\circ}$\,3639 which trace both carbon-rich polycyclic aromatic hydrocarbons (PAHs) and oxygen-rich silicate dust. With
the superior spectral coverage of SOFIA, and using a 3D photoionisation and dust radiative transfer model we prove that
the O-rich material is distributed in a shell in the outer parts of the
nebula, while the C-rich material is located in the inner parts of the
nebula. These observations combined with the model, suggest a recent change in stellar surface composition for the double chemistry in this object. This is evidence for dredge-up occurring
$\sim 10^3$\,yr ago.
\end{abstract}


\section{Introduction}

Planetary Nebulae (PNe) are the final evolutionary phase of low- and
intermediate-mass stars. The nebulae form out of the mass lost by the
star on the asymptotic giant branch (AGB), which may exceed 50\%\ of the
stellar mass. The ejecta mainly consist of gas, initially molecular and atomic
but becoming ionised by the remnant white dwarf star. Some solid particles
(`dust') condense in the ejecta.  The ejecta quickly disperses and merges with
the surrounding interstellar medium.  This recycled gas fuels most of the star
formation in late-type galaxies \citep{leiner11}. 

The chemical evolution of the Universe is driven by products of nucleosynthesis
included in stellar ejecta.  Low-mass stars are the primary source of new dust
in the Milky Way \citep{matsuura09}, and the dominant producers of carbon and nitrogen \citep{kobayashi11, henning98}.

Low- and intermediate-mass stars produce PNe with a distinct molecular and dust composition.  In
the molecular zone, the highly stable but volatile carbon monoxide (CO) molecule locks away the
less abundant element, leaving the remaining free oxygen (O) or carbon (C) to
drive the chemistry and dust formation.  If C/O$<$1 all the carbon is trapped
in CO, and the chemistry forms oxides and silicate dust, with spectral
signatures at 9.8, 18.0, 23.5, 27.5, and 33.8$\upmu$m. If C/O$>$1 then carbon
dominates the chemistry and the main observed dust features are amorphous
carbon (no spectral feature), SiC, complex hydrocarbons, including polycyclic
aromatic hydrocarbons (PAHs) (features centred at 6.2, 7.7, 8.6, and
11.3$\upmu$m), and fullerenes \citep{cami10}. Such a dichotomy between O- and C-rich PNe
is observed.
Stars begin their life with C/O$<$1 and are thus oxygen-rich.  In the
AGB phase, the dredge-up of newly synthesised carbon during a phase of helium
flashes can raise the C/O ratio above unity to form a carbon star. The details
of this process are still very uncertain \citep{karakas11, karakas14b}. 

A small fraction of PNe show both O- and C-rich features in their dust spectra,
and are therefore classified as dual chemistry objects \citep{waters98a, zijlstra91, waters98b, cohen99, cohen02, guten08, perea09, me11, me14}.  
Such objects can provide indirect evidence for
dredge-up and allow the study of the environment in which it takes place.
Two models have been proposed to explain this dual chemistry. One model requires oxygen-rich material to be present in an old and stable disk around the carbon-rich central star, and
is not related to recent mass loss \citep{zijlstra91}. There is evidence for the existence of
old, stable disks around post-AGB stars \citep{ruyter06}. The other postulates
that a recent dredge-up of carbon produced by nucleosynthesis changed the
surface chemistry of the star \citep{waters98a}. Finding objects in the process of changing from O-rich to C-rich chemistry has remained elusive \citep{zijlstra04}.

BD\,+30$^{\circ}$\,3639 (PNG 064.7+05.0) is one of the few planetary nebulae to host a Wolf-Rayet [WC] central stars. These objects show emission-line spectra similar to those of Population I Wolf-Rayet stars but have lower masses, expected for intermediate-mass stars in the post-AGB evolution \citep{leahy00}. 
Their spectra show strongly enhanced carbon and helium but have little or no hydrogen in the atmosphere. The central star of BD\,+30$^{\circ}$\,3639 (HD 184738) is a hydrogen-deficient,
carbon-rich ([WC9]) star \citep{mendez91} with an effective temperature of 55,000 K, a
luminosity of 4.25$\times$10$^3$ L$_{\odot}$, and a post-ejection mass of 0.6\,M$_{\odot}$ \citep{li02}. BD\,+30$^{\circ}$\,3639 is active in all spectral regions. While most of its energy is emitted in the mid-infrared (peaking at $\sim$30$\upmu$m), it is also a strong radio source \citep{zijlstra89}. It also has a molecular envelope with a mass of 0.016\,M$_{\odot}$ estimated from millimetre rotational lines of CO \citep{bachiller91}.

\section{Observations}

Using the Faint Object InfraRed CAmera (FORCAST) for the Stratospheric
Observatory for Infrared Astronomy (SOFIA) Telescope \citep{adams10} we obtained images of BD\,+30$^{\circ}$\,3639. The
observations cover the wavelength ranges of both the PAHs and crystalline
silicate features. SOFIA uniquely allows
for the imaging of the long wavelength emission from crystalline silicate
features at high angular resolution.

The FORCAST instrument has a short wavelength camera (SWC) that operates from 5-25$\upmu$m and a long wavelength camera (LWC) that operates from 25-40$\upmu$m, with several filters available in both cameras. FORCAST samples at 0.75$^{\prime\prime}$/pixel, giving a 3.2$^{\prime}\times$ 3.2$^{\prime}$ instantaneous field-of-view. Five different filters were used: FOR\_F064 (centred at 6.4$\upmu$m with a $\Delta\lambda$=0.14$\upmu$m), FOR\_F077 (centred at 7.7$\upmu$m with a $\Delta\lambda$=0.47$\upmu$m), FOR\_F111 (centred at 11.1$\upmu$m with a $\Delta\lambda$=0.95$\upmu$m), FOR\_F113 (centred at 11.3$\upmu$m with a $\Delta\lambda$=0.24$\upmu$m), and FOR\_F336 (centred at 33.6$\upmu$m with a $\Delta\lambda$=1.9$\upmu$m). We used the symmetric nod-match-chop (NMC) imaging mode, the chop is symmetric about the optical axis of the telescope with one of the two chop positions centred on the target. The nod throw is oriented 180$^{\circ}$ from the chop, such that when the telescope nods, the source is located in the opposite chop position. 

\section{Results}

In Figure 1 we present the images of BD\,+30$^{\circ}$\,3639 obtained using the filters
centred at 6.4, 7.7, 11.1, 11.3, and 33.6$\upmu$m. Each image has an over-plotted blue circle of 5$^{\prime\prime}$ in radius. To study the spatial
extension of the different dust features in the PN, we performed a radial cut
along the major axis of the nebula (PA=135$^{\circ}$) for all the SOFIA
images (the white line in the top left panel represents the PA of this cut). Figure 2 shows these cuts, the top plot shows the cut made using
the PAH filters (centred at 6.4$\upmu$m, 7.7$\upmu$m, 11.1$\upmu$m, 11.3$\upmu$m). The extent of the shells in the PAHs filters is very similar. The cuts of the 7.7 and the
11.3$\upmu$m images show a double peaked emission, as expected for a shell
surrounding a hollow interior. The measured full width at half maximum (FWHM) is 8$^{\prime\prime}$ at these wavelengths. The middle image shows the PAHs filters and the silicates filter (centred at 33.6$\upmu$m).   The emission at 33.6$\upmu$m is 1.5 times more extended than the C-rich material, with a FWHM of
12$^{\prime\prime}$. The bottom plot shows the same radial cuts done in all the filters with the normalised flux. 

Figure 3 shows the Infrared Space Observatory (ISO) Short Wavelength Spectrometer (SWS) spectrum of BD\,+30$^{\circ}$\,3639 (reproduced from \citealt{sloan03}), with the bandpass of the SOFIA filters overlaid by a dashed line of different colours depending on the filter. Dark-blue represents the bandpass for the FOR\_F064 filter, the FOR\_F077 filter bandpass is in purple, brown for the FOR\_F111 filter, green for the FOR\_F113 filter, and light-blue for the FOR\_F336 filter. Using the ISO spectrum we can clearly see the PAHs and the silicates features covered by each filter. We can see the main PAHs (6.25$\upmu$m, 7.7$\upmu$m, and 11.3$\upmu$m) as well as the crystalline silicates (33.6$\upmu$m) features.

Previous observations of BD\,+30$^{\circ}$\,3639 show a shell-like structure with a
radius of 5$^{\prime\prime}\times$4$^{\prime\prime}$ in the visible \citep{li02} and
6$^{\prime\prime}\times$5$^{\prime\prime}$ in radio \citep{bryce97}, while a much larger
extension of the faintest part of the nebula is detected in the near-infrared
\citep{phillips07}. Previous mid-infrared imaging observations of BD\,+30$^{\circ}$\,3639 with a
spatial resolution of 1.5$^{\prime\prime}$ found that
the 8.6 and 11.3$\upmu$m (PAHs) bands are slightly more extended than the
continuum emission \citep{bentley84}.  Higher spatial resolution imaging and
spectroscopic observations of BD\,+30$^{\circ}$\,3639, taken with the Subaru telescope,
confirmed that the PAH emission bands are more extended than the dust continuum
emission, but show a similar extent to the [Ne II] 12.8$\upmu$m emission
\citep{matsumoto08}. These observations also show evidence for a 10$\upmu$m absorption feature, attributed to silicates. 

Until SOFIA, it was not possible to map the silicate emission features in this object, because the features emit at longer infrared wavelengths where no instrument was able to offer the necessary combination of wavelength coverage, sensitivity, and spatial resolution.  
Using the expansion velocity of the silicate
shell of 10 km/s, and a distance of 1.2\,kpc \citep{li02}, the expansion rate of the shell is
1.76$\pm$0.25 mas\,yr$^{-1}$ in radius. Using the images we compared the size of the C-rich to the O-rich shell. For the C-rich shell we averaged all the images of the PAHs filters (6.4 to 11.3$\upmu$m), for the O-rich shell we used the image of the silicates filter (33.6$\upmu$m). We measure the size (radius) of the C-rich and the O-rich shells
to be 5$\pm$0.75$^{\prime\prime}$ and 7.5$\pm$0.75$^{\prime\prime}$,
respectively. These correspond to dynamical ages of 2,800$\pm$580 yr for the
C-rich shell and 4,300$\pm$740 yr for its O-rich counterpart. This means that
the transition of the star from O- to C-rich took place within a window of
1,500$\pm$940 yr. These values are entirely consistent with the post-AGB age
of the central star determined by comparison of its luminosity and temperature
to evolutionary tracks \citep{vassi93, blo01}. 

\begin{figure}
\centering
\makebox{
\includegraphics[width=4.25cm, height=2.9cm]{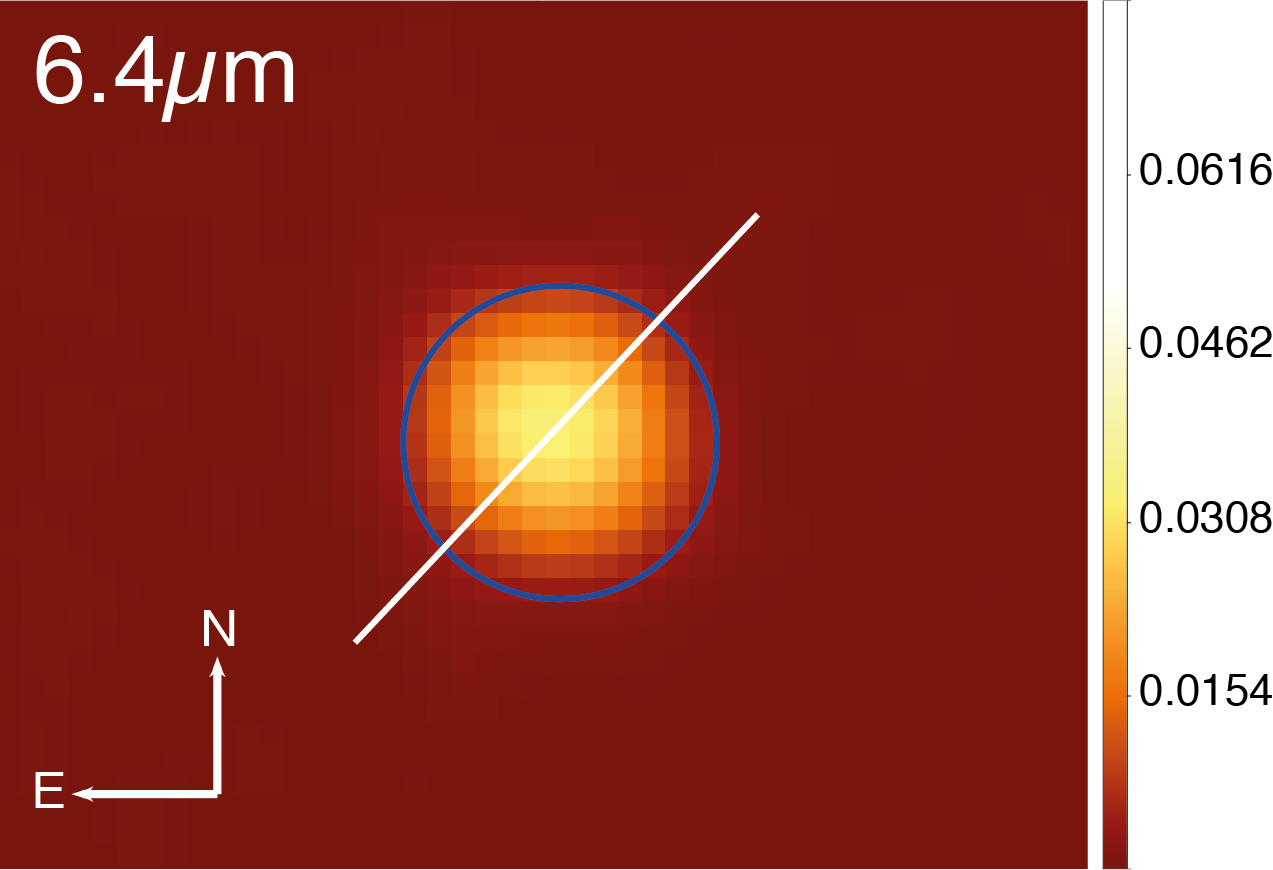}
\includegraphics[width=4.25cm, height=2.9cm]{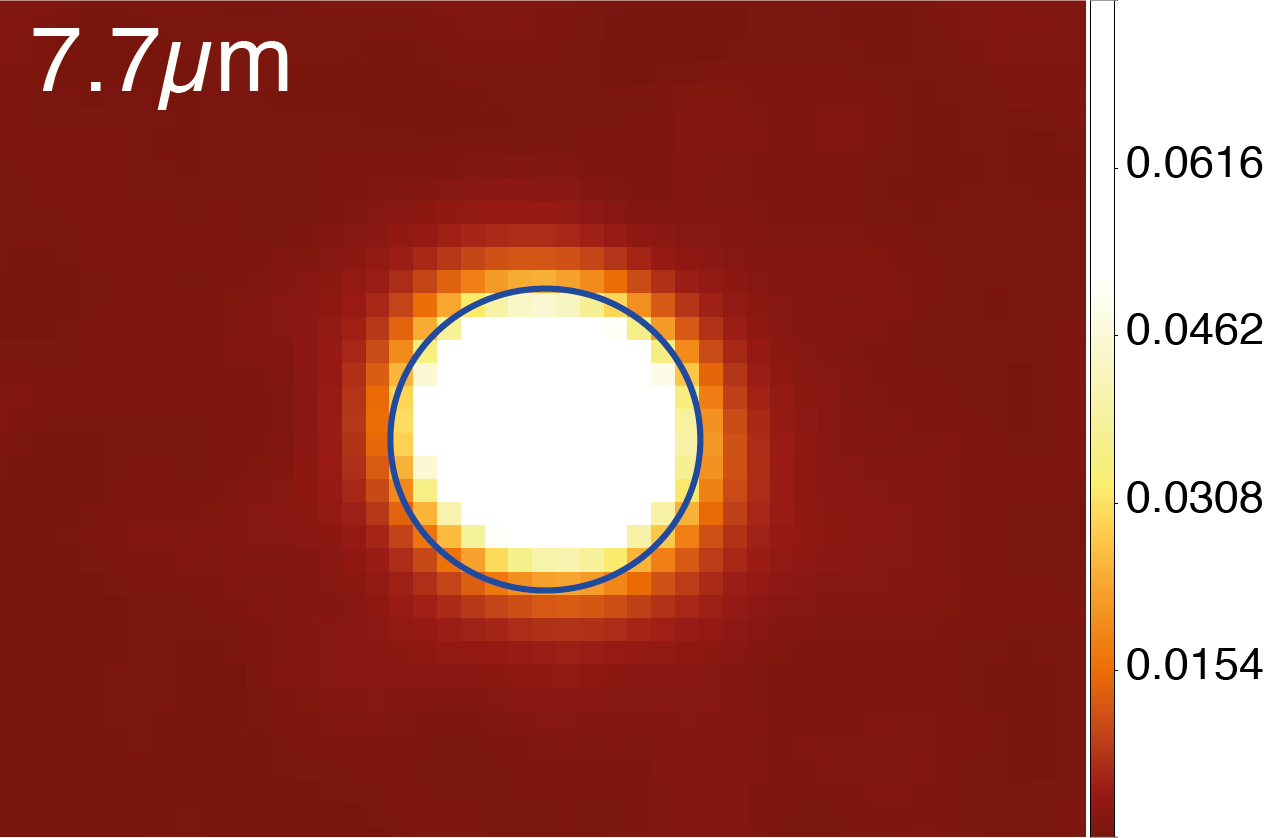}}
\vspace{0.10cm}
\vspace{0.0cm}
\makebox{
\includegraphics[width=4.25cm, height=2.9cm]{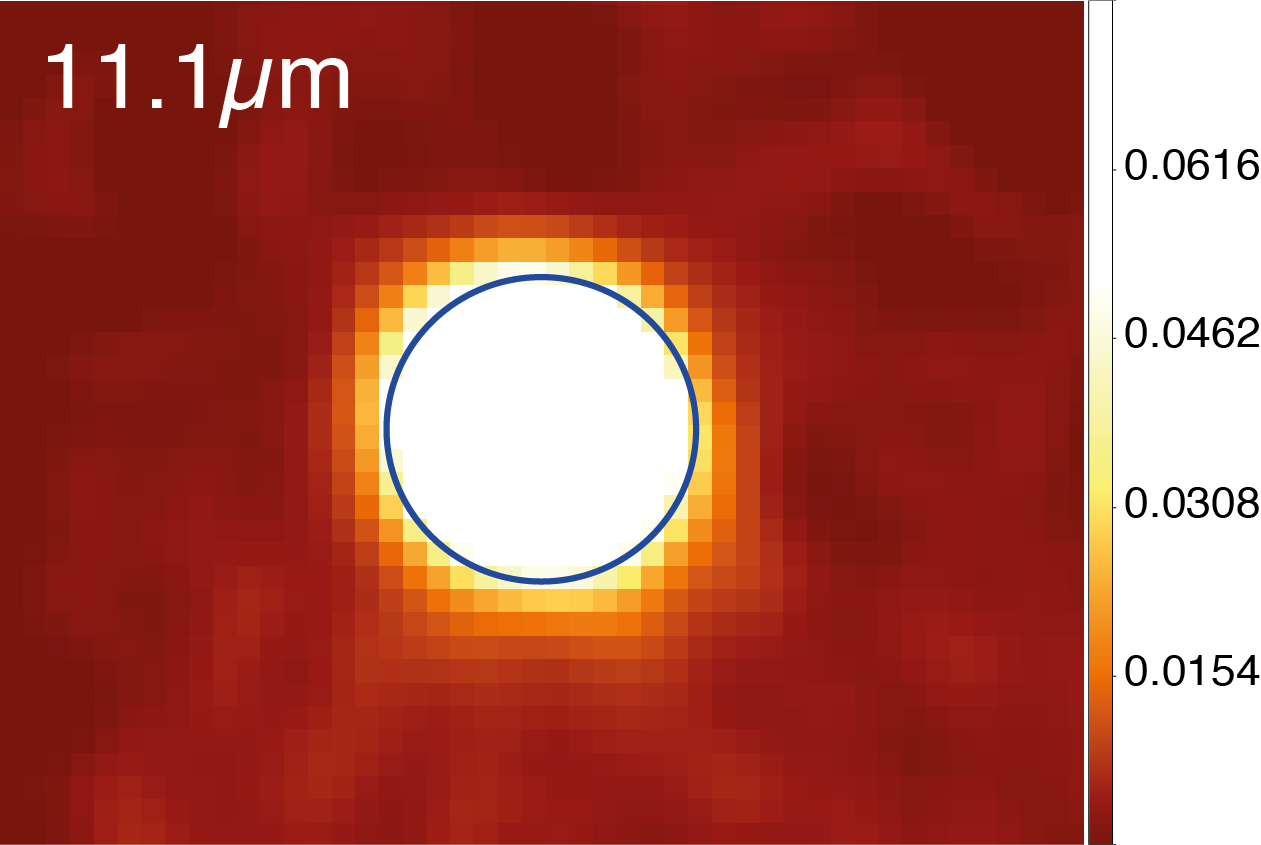}
\includegraphics[width=4.25cm, height=2.9cm]{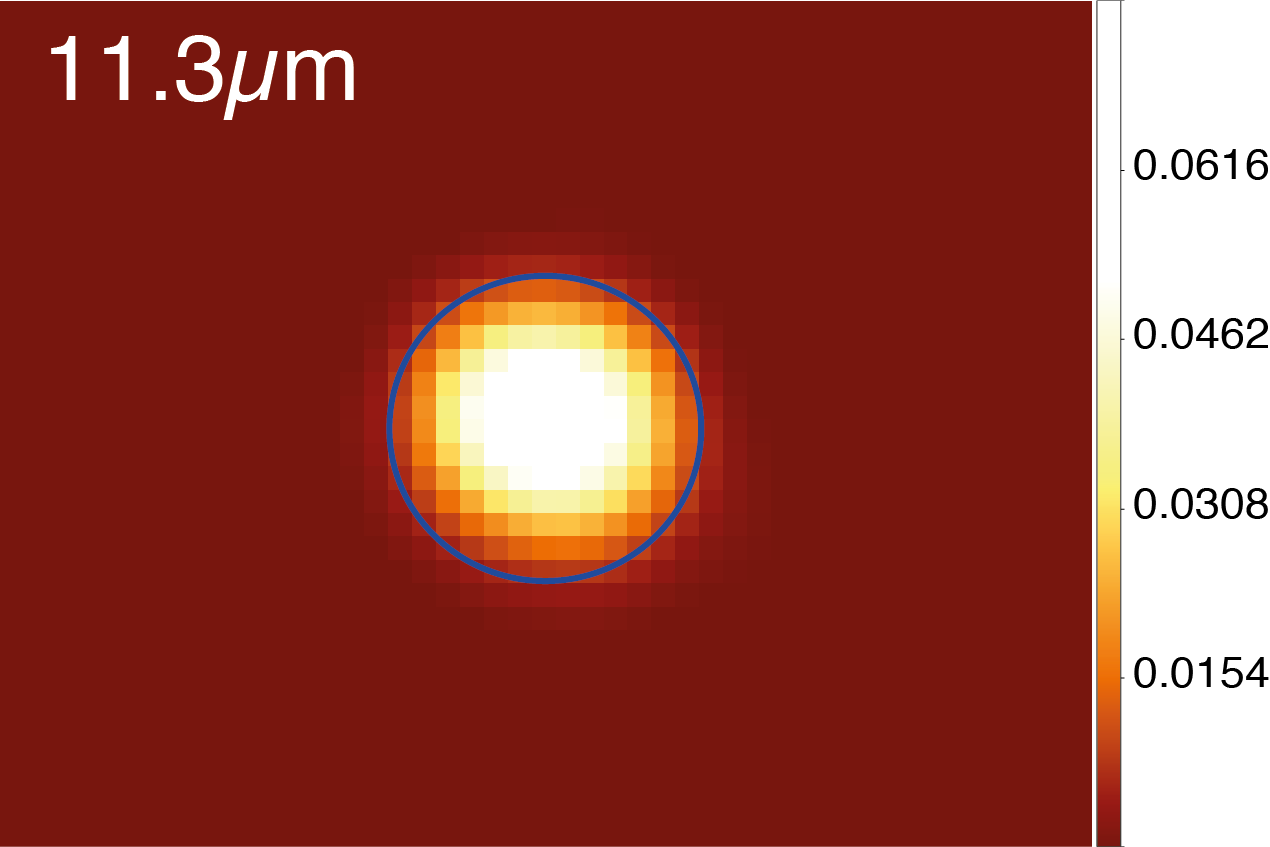}}
\\
\vspace{0.4cm}
\includegraphics[width=4.25cm, height=2.9cm]{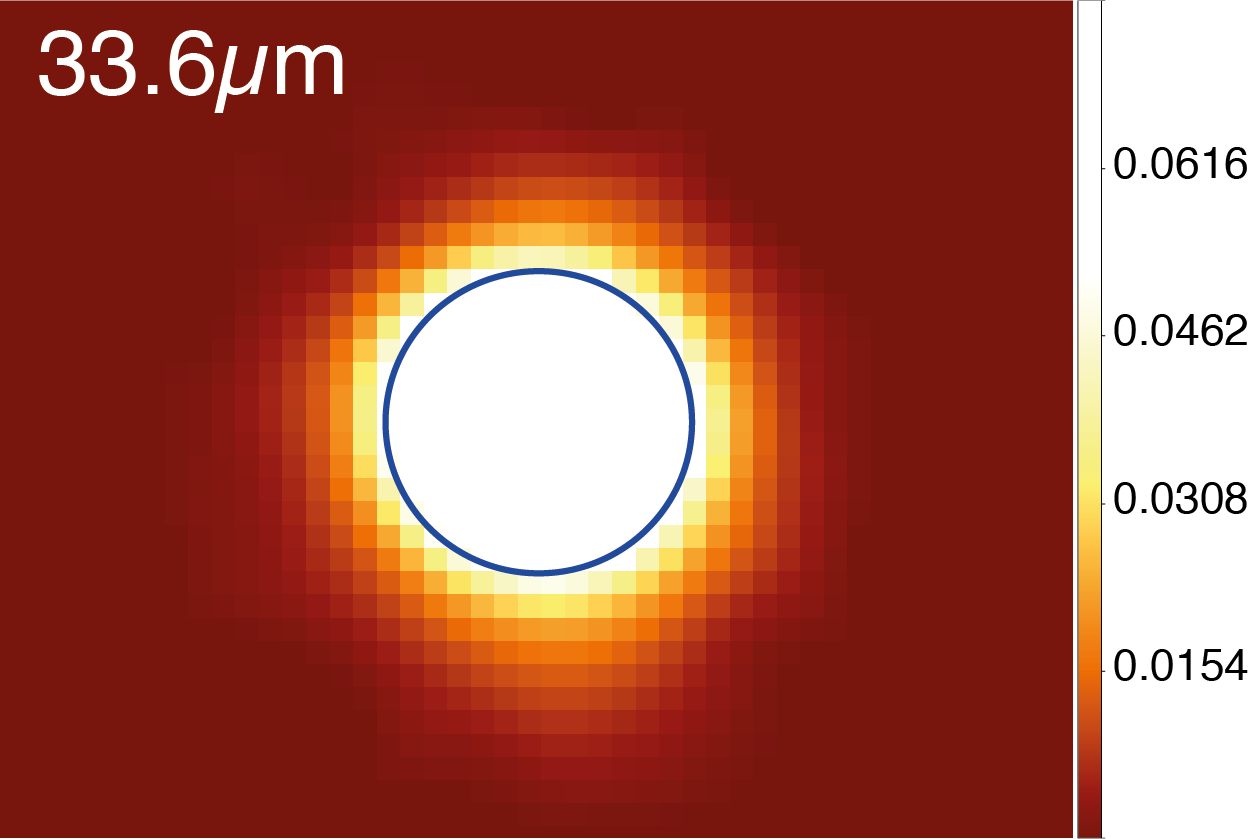}
\caption{SOFIA images of BD\,+30$^{\circ}$\,3639 using the 6.4$\upmu$m, 7.7$\upmu$m, 11.1$\upmu$m, 11.3$\upmu$m, and 33.6$\upmu$m filters. To compare the size of each image, we have overplotted a blue circle of 5$^{\prime\prime}$ radius in all of them. The white line in the top left panel represents the PA of the cut made to all the images.}
\end{figure}    

\begin{figure}
\centering
\includegraphics[width=5.8cm, height=4.8cm]{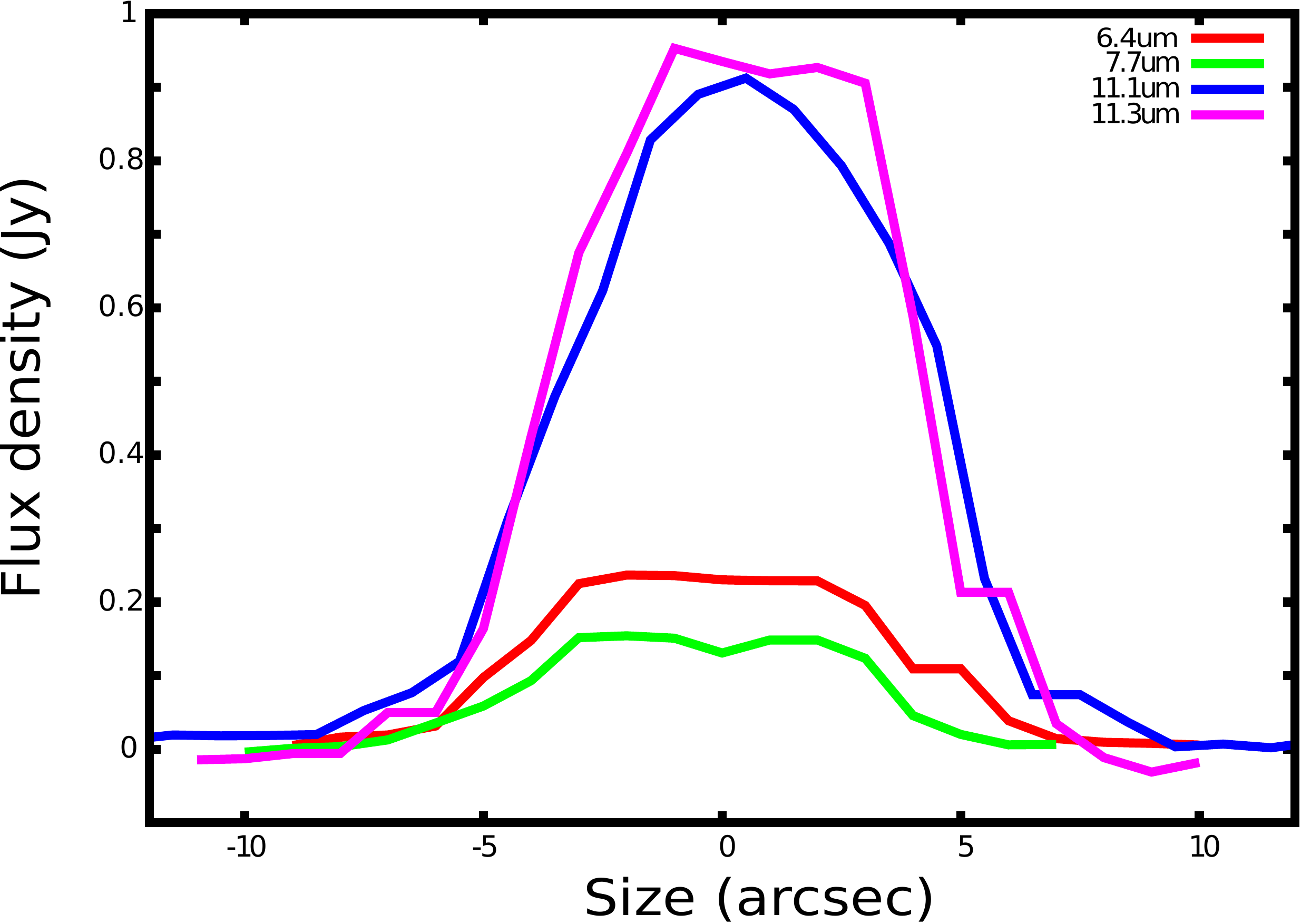}
\includegraphics[width=5.8cm, height=4.8cm]{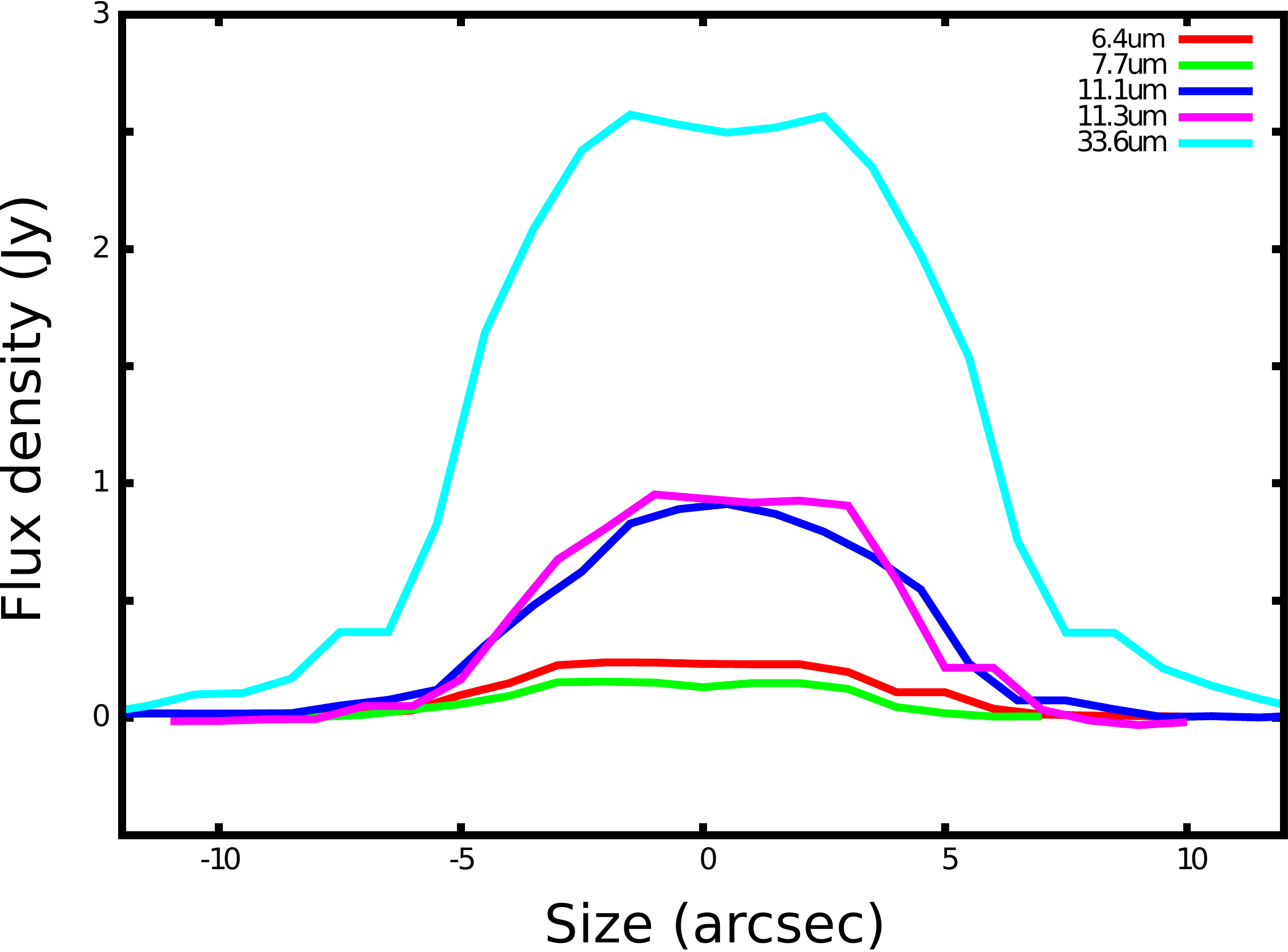}
\includegraphics[width=5.9cm, height=4.8cm]{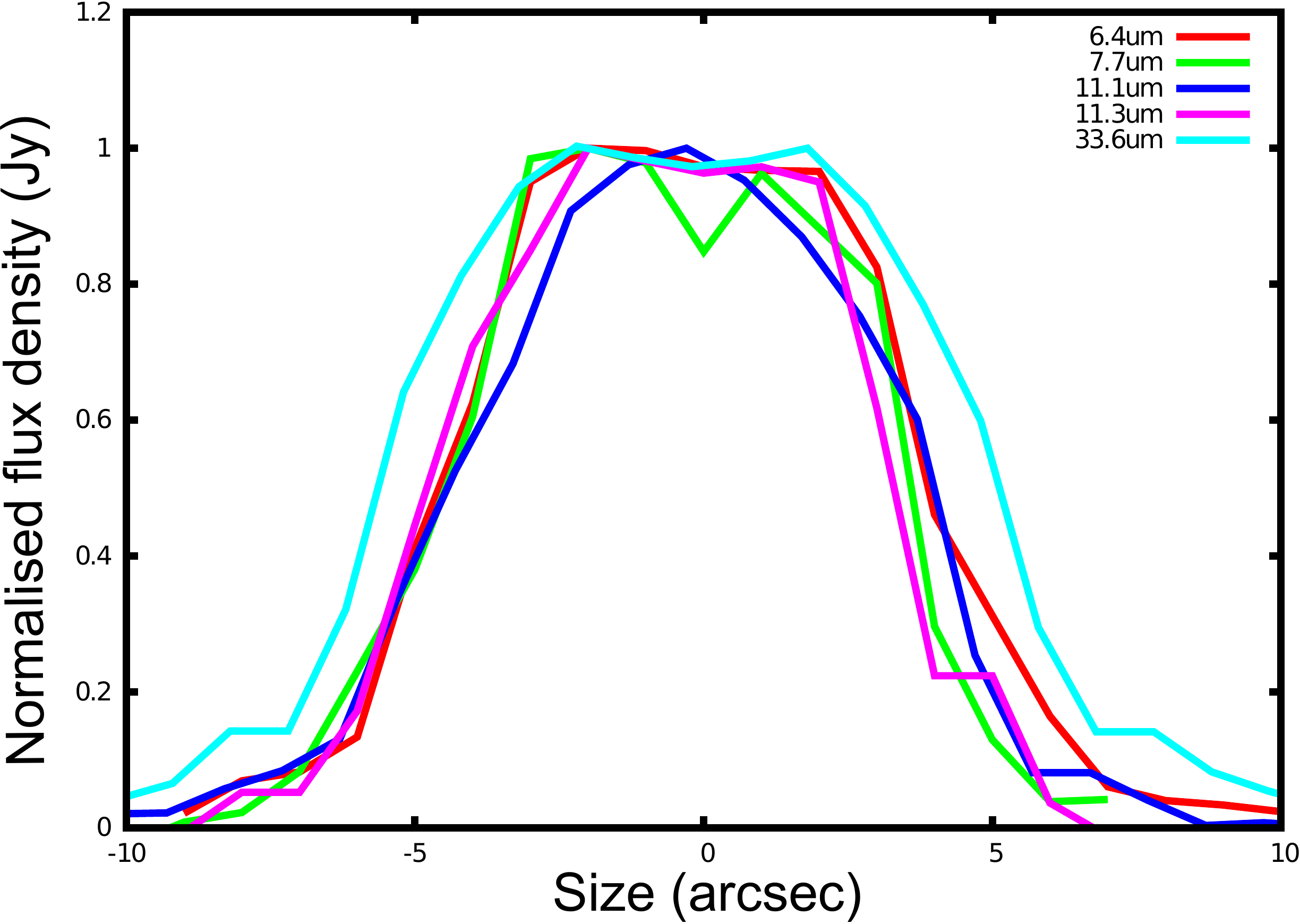}
\caption{Radial cuts along the major axis of BD\,+30$^{\circ}$\,3639, along a PA of 135$^{\circ}$ for all the SOFIA images: Top image shows the PAHs filters only (6.4$\upmu$m, 7.7$\upmu$m, 11.1$\upmu$m, 11.3$\upmu$m). The middle image shows the PAHs filters and the silicates filter (33.6$\upmu$m), while the bottom image shows all the filters with the flux normalised.}
\end{figure}    

\begin{figure}
\includegraphics[width=8.5cm, height=7cm]{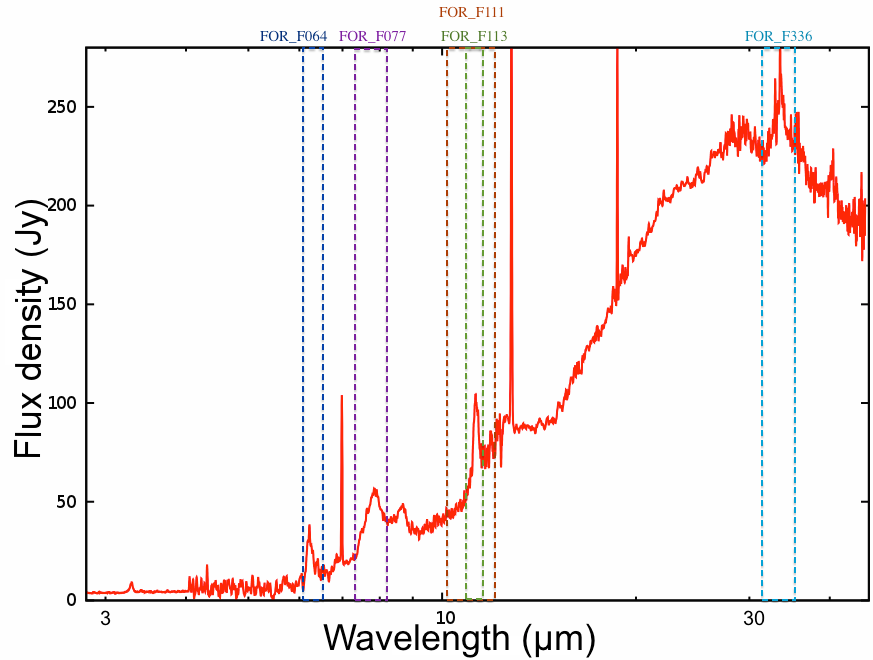}
\caption{ISO spectrum of BD\,+30$^{\circ}$\,3639. The dashed lines are used to show the position and width of the SOFIA filters used on this paper. Dark-blue represents the bandpass for the FOR\_F064 filter, the FOR\_F077 filter bandpass is in purple, brown for the FOR\_F111 filter, green for the FOR\_F113 filter, and light-blue for the FOR\_F336 filter.}
\end{figure}    

\section{Analysis \& Discussion}

In Figure 1 and 2, we can see the different extensions of the two major components presented here, PAHs and silicates. It is important to note that no continuum filter observations were made, as such, the dust continuum contribution is not subtracted from any of the presented observations. Therefore, one must consider the level of continuum contamination in each filter before drawing conclusions as to the relative distribution of PAHs and silicates. \\
In the case of PAHs, from Figure 3, we can see that the FOR\_F064 filter has a negligible contamination by dust continuum and thus traces the PAHs emission. From Figure 2 we can see that the emission in this filter has the same extension as the other PAHs filters, as such, we can assume that the measurements made in the PAHs filters are tracing the PAH emission, rather than just extended dust continuum.\\
Unfortunately, the observation made in the filter encompassing the silicates emission, the FOR\_F336 filter, includes a significant contribution of dust continuum, as shown in Figure 3. Based on the images alone, we are unable to determine the spatial distribution of the silicates with respect to the dust continuum in the nebula. In order to resolve this degeneracy, a radiative transfer model of the nebula was created.

We constructed simple 3D photoionisation and dust radiative transfer models using the {\sc mocassin} code \citep{ercolano03, ercolano05, ercolano08}. The model consists of neutral PAHs and graphitic carbon (optical constants from \citealt{li01}), and amorphous silicates (optical constants from \citealt{laor93}) and crystalline fosterite (optical constants from \citealt{jager98}). In both the dust density is proportional to r$^{-2}$ and the grain size distribution is standard Mathis, Rumpl, \& Nordsieck (1977 [MRN]). We adopted a distance of 1.2\,kpc, and a central star luminosity and temperature of 4.25$\times$10$^3$\,L$_{\odot}$ and 55,000 K. We varied the shell
radii, the position of the silicates and carbon, and the mass of dust in each shell to obtain a good fit to the data. For the model to fit the ISO spectrum and the SOFIA data points, the carbon material has to be in the inner parts, while the silicates need to be in the outer parts of the nebula. If silicate dust is present in the inner warmer regions of the shell, emission features at 10 and 18$\upmu$m are predicted, opposite to what it is observed. The total mass is constrained to within 10\% by the models. 
In the best fitting model (Figure 4), the carbon shell extends from 1.5 to 5.5$^{\prime\prime}$ on the plane of the sky and contains 6.6$\times$10$^{-5}$\,M$_{\odot}$ of dust, while the silicate shell extends from 5.5 to 7.5$^{\prime\prime}$ and has a mass of 8.1$\times$10$^{-5}$\,M$_{\odot}$. The carbon dust mass loss rate, assuming the switch happened 2,800\,yr ago, would be 2.4$\times$10$^{-8}$\,M$_{\odot}$\,yr$^{-1}$, so a total mass loss rate of 7.0$\times$10$^{-6}$\,M$_{\odot}$\,yr$^{-1}$ if the gas-to-dust ratio is 300. The silicate dust mass loss rate would be 5.4$\times$10$^{-8}$\,M$_{\odot}$\,yr$^{-1}$ for a total mass loss rate of 1.6$\times$10$^{-5}$\,M$_{\odot}$\,yr$^{-1}$.\\

\begin{figure}
\label{cuts}
\centering
\includegraphics[width=8.5cm, height=7cm]{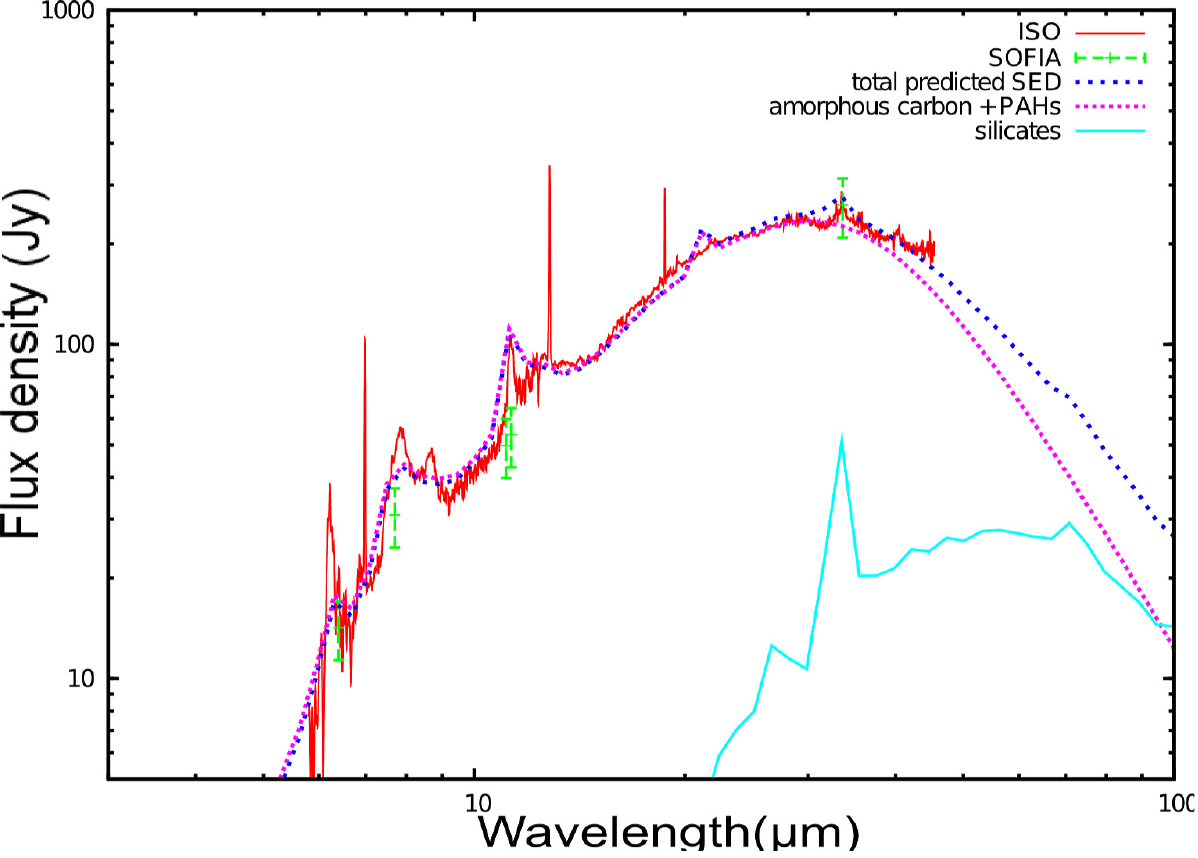}
\caption{3D model using {\sc mocassin}. The blue dotted line is the total SED given by the model, in cyan we show the contribution of the O-rich dust and in pink the C-rich dust. The green crosses are the SOFIA fluxes (assuming a 20\% uncertainty), and the red line is the ISO spectrum over-plotted.}
\end{figure}

The observed C/O ratio in the ionized region is $1.59$ \citep{bernard03}. Assuming
instant, homogeneous mixing, and a solar O abundance, an initial C/O ratio of 0.8, and an envelope mass of 0.01\,M$_\odot$, the numbers suggest that the thermal pulse dredged-up $4
\times 10^{-5}\,\rm M_\odot$ of carbon, this is almost the same value we obtain. However, the current wind from the central star has $\rm C/O = 30\pm15$ \citep{yu09}, so that even if the
mass-loss rates are low, the current wind can still add significantly to the
carbon budget. The range of C/O and the somewhat low carbon dust mass may
suggest that the carbon dredge-up was not instantaneous but that the C/O
increased while the mass loss continued.

Independently, a stellar evolutionary model was constructed with an initial
stellar mass of 1.5\,M$_\odot$ and solar metallicity, fairly typical values for
a Galactic Disk PN. This model (Figure 5) derives a peak
mass-loss rate at the point where the model star becomes C-rich of
$\sim$1$\times$10$^{-5}$\,M$_{\odot}$\,yr$^{-1}$ and a final central star remnant mass of 0.62\,M$_{\odot}$ \citep{karakas14}, closely matching those values determined from the
observations and photoionisation modelling.  The model shows that the typical
interpulse period near the end of the star's life is $\sim$1$\times$10$^{5}$
years, while the total duration of a thermal pulse, a third dredge-up, and
relaxation back to H-shell burning is of the order of $\sim$1000 years or
more. The numbers fit the duration of the change from oxygen-rich to
carbon-rich mass loss observed with SOFIA, and confirm that this is the
likely explanation. At the last thermal pulse, $4.0\times 10^{-3}\, \rm
M_\odot $ is dredged-up to the envelope. About $ 1 \times 10^{-3}\, \rm
M_\odot $ of this is $^{12}$C. This is higher than derived from the
observations.  Incomplete mixing \citep{hajduk05} can therefore not be ruled out.

\begin{figure}
\label{cuts}
\centering
\includegraphics[width=8.5cm, height=7cm]{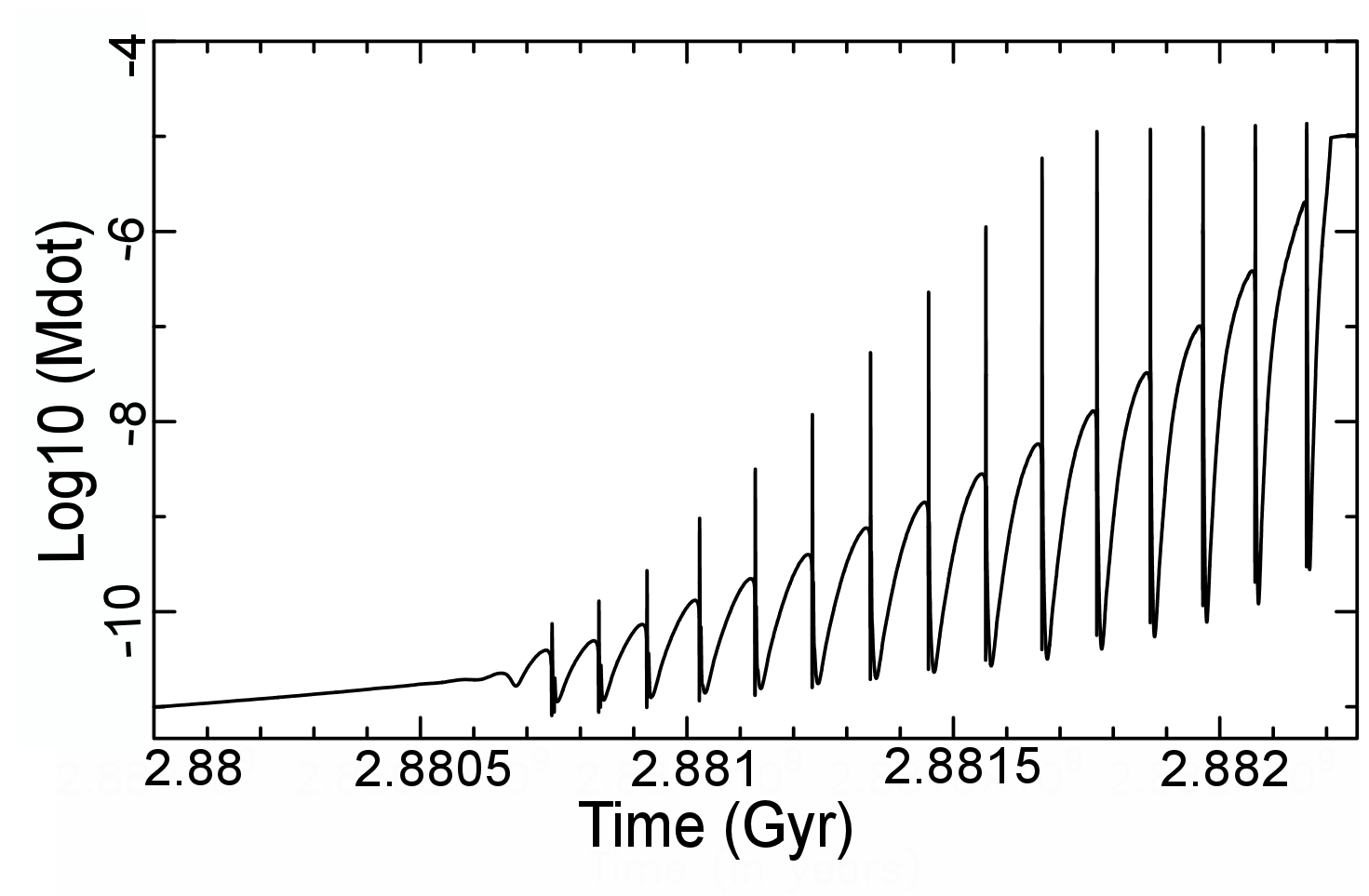}
\caption{Evolution of the mass-loss rate with time, as estimated from our stellar evolution model. We used an initial mass of the star of 1.5\,M$_{\odot}$, and a solar metallicity (Z = 0.014). The peak mass-loss rate is $\sim$1$\times$10$^{-5}$\,M$_{\odot}$\,yr$^{-1}$ and the final mass of the star is 0.62\,M$_{\odot}$, which closely matches the value calculated from the observations.}
\end{figure}

The observations and modelling suggest that the thermal pulse took place while the
envelope mass was already very low. This is consistent with an AGB final (or
fatal) thermal pulse (AFTP) occurring immediately before the star moves off
the AGB. An AFTP can lead to both a considerable enrichment with carbon and
oxygen and to the dilution of hydrogen, and also explains the current
hydrogen-poor nature of the current central star \citep{blo01}.

The short duration of this phase compared to the long interpulse period
suggests that such observations of carbon dredge-up will remain
uncommon. BD+30\,3639 provides a rare glimpse into a phase of stellar
evolution which is crucial to the origin of carbon in the Universe.

\section*{Acknowledgments}
This work was co-funded under the Marie Curie Actions of the European Commission
(FP7-COFUND). AIK was supported through an Australian Research Council Future Fellowship (FT110100475). LGR thanks Thomas Kr$\rm{\ddot{u}}$hler for all the proactivity, the high level suggestions, the world class comments, and the highly efficient discussion on this paper. The authors thank the important comments of the referee (Jeronimo Bernard-Salas) that made the paper clearer and more precise.

\bsp

\label{lastpage}


\newpage
\begin{thebibliography}{99}
\bibitem[\protect\citeauthoryear{Adams et al.}{2010}]{adams10} Adams et al., 2010, Proc. SPIE 7735, 77351U  
\bibitem[\protect\citeauthoryear{Bachiller et al.}{1991}]{bachiller91} Bachiller et al., 1991, A\&A, 247, 525  
\bibitem[\protect\citeauthoryear{Bentley et al.}{1984}]{bentley84} Bentley et al., 1984, ApJ, 278, 665  
\bibitem[\protect\citeauthoryear{Bernard-Salas et al.}{2003}]{bernard03} Bernard-Salas et al., 2003, A\&A, 406, 165 
\bibitem[\protect\citeauthoryear{Bl$\ddot{\rm o}$cker}{2001}]{blo01} Bl$\ddot{\rm o}$cker, 2001, Ap\&SS, 275, 1 
\bibitem[\protect\citeauthoryear{Bryce et al.}{1997}]{bryce97} Bryce et al., 1997, MNRAS, 284, 815
\bibitem[\protect\citeauthoryear{Cami et al.}{2010}]{cami10} Cami et al., 2010, {\it Science}, 329, 1180 
\bibitem[\protect\citeauthoryear{Cohen et al.}{1999}]{cohen99} Cohen et al., 1999, ApJ, 513, L135  
\bibitem[\protect\citeauthoryear{Cohen et al.}{2002}]{cohen02} Cohen et al., 2002, MNRAS, 332, 879  
\bibitem[\protect\citeauthoryear{de Ruyter et al.}{2006}]{ruyter06} de Ruyter et al., 2006, A\&A, 448, 641  
\bibitem[\protect\citeauthoryear{Ercolano et al.}{2003}]{ercolano03} Ercolano et al., 2003, MNRAS, 340, 1136 
\bibitem[\protect\citeauthoryear{Ercolano et al.}{2005}]{ercolano05} Ercolano et al., 2005, MNRAS, 362, 1038 
\bibitem[\protect\citeauthoryear{Ercolano et al.}{2008}]{ercolano08} Ercolano et al., 2008, ApJS, 175, 534 
\bibitem[\protect\citeauthoryear{Gutenkunst et al.}{2008}]{guten08} Gutenkunst et al., 2008, ApJ, 680, 1206
\bibitem[\protect\citeauthoryear{Guzman-Ramirez et al.}{2011}]{me11} Guzman-Ramirez et al., 2011, MNRAS, 414, 1667 
\bibitem[\protect\citeauthoryear{Guzman-Ramirez et al.}{2014}]{me14} Guzman-Ramirez et al., 2014, MNRAS, 441, 364 
\bibitem[\protect\citeauthoryear{Hajduk et al.}{2005}]{hajduk05} Hajduk et al., 2005, {\it Science}, 308, 231 
\bibitem[\protect\citeauthoryear{Henning \& Salama}{1998}]{henning98} Henning \& Salama, 1998, {\it Science}, 282, 2204 
\bibitem[\protect\citeauthoryear{Jager et al.}{1998}]{jager98} Jager et al., 1998, A\&A, 332, 291
\bibitem[\protect\citeauthoryear{Karakas}{2011}]{karakas11} Karakas, 2011, ASPC, 445, 3 
\bibitem[\protect\citeauthoryear{Karakas \& Lattanzio}{2014}]{karakas14b} Karakas \& Lattanzio, PASA, 2014, 31, 30
\bibitem[\protect\citeauthoryear{Karakas}{2014}]{karakas14} Karakas, 2014, MNRAS, 445, 374 
\bibitem[\protect\citeauthoryear{Kausch et al.}{2014}]{kaush14} Kausch et al., 2014, submitted to A\&A 
\bibitem[\protect\citeauthoryear{Kobayashi et al.}{2011}]{kobayashi11} Kobayashi et al., 2011, MNRAS, 414, 3231  
\bibitem[\protect\citeauthoryear{Laor \& Draine}{1993}]{laor93} Laor \& Draine, 1993, ApJ, 402, L441 
\bibitem[\protect\citeauthoryear{Leahy et al.}{2000}]{leahy00} Leahy et al., 2000, ApJ, 540, 442
\bibitem[\protect\citeauthoryear{Leitner \& Kravtsov}{2011}]{leiner11} Leitner \& Kravtsov, 2011, ApJ, 734, 48 
\bibitem[\protect\citeauthoryear{Li \& Draine}{2001}]{li01} Li \& Draine, 2001, ApJ, 550, L213 
\bibitem[\protect\citeauthoryear{Li et al.}{2002}]{li02} Li et al., 2002, AJ, 123, 2676 
\bibitem[\protect\citeauthoryear{Mathis et al.}{1977}]{mathis77} Mathis et al., 1977, ApJ, 217, 425
\bibitem[\protect\citeauthoryear{Matsuura et al.}{2009}]{matsuura09} Matsuura et al., 2009, MNRAS, 396, 918  
\bibitem[\protect\citeauthoryear{Matsumoto et al.}{2008}]{matsumoto08} Matsumoto et al., 2008, ApJ, 667, 1120 
\bibitem[\protect\citeauthoryear{M\'endez}{1991}]{mendez91} M\'endez, 1991, IAU Symp. 145, 375
\bibitem[\protect\citeauthoryear{Perea-Calderon et al.}{2009}]{perea09} Perea-Calder{\'o}n et al., 2009, A\&A, 495, L5 
\bibitem[\protect\citeauthoryear{Phillips \& Ramos-Larios}{2007}]{phillips07} Phillips \& Ramos-Larios, 2007, AJ, 133, 347  
\bibitem[\protect\citeauthoryear{Siebenmorgen et al.}{1994}]{siebe94} Siebenmorgen et al., 1994, MNARS, 271, 449 
\bibitem[\protect\citeauthoryear{Sloan et al.}{2003}]{sloan03} Sloan et al., 2003, ApJS, 147, 379
\bibitem[\protect\citeauthoryear{Smette et al.}{2015}]{smette15} Smette et al., 2015, A\&A, in press
\bibitem[\protect\citeauthoryear{Vassiliadis \& Wood}{1993}]{vassi93} Vassiliadis \& Wood, 1993, ApJ, 413, 641 
\bibitem[\protect\citeauthoryear{Waters et al.}{1998}]{waters98a} Waters et al., 1998, A\&A, 331, L61 
\bibitem[\protect\citeauthoryear{Waters et al.}{1998}]{waters98b} Waters et al., 1998, Nature, 391, 868 
\bibitem[\protect\citeauthoryear{Yu et al.}{2009}]{yu09} Yu et al., 2009, ApJ, 690, 440  
\bibitem[\protect\citeauthoryear{Zijlstra et al.}{1989}]{zijlstra89} Zijlstra et al., 1989, A\&AS, 79, 329
\bibitem[\protect\citeauthoryear{Zijlstra et al.}{1991}]{zijlstra91} Zijlstra et al., 1991, A\&A, 243, L9 
\bibitem[\protect\citeauthoryear{Zijlstra et al.}{2004}]{zijlstra04} Zijlstra et al., 2004, MNRAS, 352, 325 
  
\end{thebibliography}
\end{document}